\documentclass[reprint,twocolumn,showpacs,preprintnumbers,showkeys,amsmath,amssymb,pra]{revtex4-1}
\usepackage{graphicx}% Include figure files
\usepackage{dcolumn}% Align table columns on decimal point
\usepackage{bm}% bold math

\usepackage[T1]{fontenc} %inne kodowanie (to wlaczamy ja usuniemy [MeX]{polski}
\newcommand{\wymiar}{0.45\textwidth}
\usepackage{hyperref}

\newcommand{\wyroz}[1]{#1}

\begin{document}

\preprint{Submitted to: ACTA PHYSICA POLONICA A}

\title{Superconductivity, metastability and magnetic field induced phase separation\\ in the atomic limit  of the Penson-Kolb-Hubbard model}% Force line breaks with \\

\author{Konrad Jerzy Kapcia}%
% \altaffiliation[Also at ]{Physics Department, XYZ University.}%Lines break automatically or can be forced with \\
    \email{ e-mail: konrad.kapcia@amu.edu.pl}
%\author{Stanis\l{}aw Robaszkiewicz}%
\affiliation{Electron States of Solids Division, Faculty of Physics, Adam Mickiewicz University in Pozna\'n, Umultowska 85, 61-614 Pozna\'n, Poland
}%

\date{October 6, 2014}% It is always \today, today, but any date may be explicitly specified

\begin{abstract}
We present the analysis of paramagnetic effects of magnetic field ($B$) (Zeeman term)
in the zero-bandwidth limit of the extended Hubbard model
for  arbitrary chemical potential $\mu$ and electron density $n$.
The effective Hamiltonian considered consists of  the  on-site interaction $U$ and the  intersite charge exchange term $I$, determining the hopping of electron pairs between nearest-neighbour sites.
The model has been analyzed within the variational approach, which treats the on-site interaction term exactly and the intersite interactions within the mean-field approximation (rigorous in the limit of infinite dimensions \mbox{$d\rightarrow+\infty$}).
In this report we focus on metastable phases as well as phase separated (PS) states involving superconducting (SS) and nonordered (NO) phases and determine their ranges of occurrence for \mbox{$U/I_0=1.05$} (\mbox{$I_0=zI$}) in the presence of magnetic field \mbox{$B\neq0$}. Our investigations of the general case for arbitrary $U/I_0$ show that, depending on the values of interaction parameters (for fixed $n$), the PS state can occur in higher fields than the homogeneous SS phase (field-induced PS). Moreover, a first-order SS--NO transition occurs between  metastable phases and these metastable phases can exist inside the regions of the PS state stability. Such behaviour is associated with the presence of tricritical line on the phase diagrams of the system.

\end{abstract}

% PACS, the Physics and Astronomy
                             % Classification Scheme.
%Use showkeys class option if keyword
                              %display desired

\maketitle

%===================================================================================================

\section{General formulation}\label{sec:intro}

The purpose of the present work is the analysis of paramagnetic effects of magnetic field (Zeeman term) on metastability in the zero-bandwidth limit of the extended Hubbard model with pair hopping interaction \cite{RP1993,KRM2012,KR2013,K2014,N1} (i.e. the \mbox{$t=0$} limit of the so-called Penson-Kolb-Hubbard (PKH) model).

The PKH model is one of the conceptually simplest effective models for studying superconductivity of the narrow band systems with short-range, almost unretarded pairing \cite{N1,HD1993,RB1999,MM2004,PMM2009,PMM2011,RCz2001,CzR2004,JM1997,JKS2001,N2}. The model includes a nonlocal pairing mechanism that is distinct from on-site interaction in the attractive Hubbard model and that is the driving force of pair formation and also of their condensation.

Because of the complexity of the PKH model there are no exact solutions for that model. In this paper we present the \mbox{$d\rightarrow+\infty$} exact results for the PKH model with \mbox{$t=0$}. We extend our investigations of the model  to the case of finite external magnetic field $B$ and concentrate on metastable phases and phase separations in the case of \mbox{$B\neq0$}.

Our starting point is the model with the hamiltonian given by
\begin{eqnarray}
\hat{H} & = & U\sum_{i}{\hat{n}_{i\uparrow}\hat{n}_{i\downarrow}}- I\sum_{\langle i,j\rangle}{(\hat{\rho}_i^+\hat{\rho}_j^- + \hat{\rho}_j^+\hat{\rho}_i^- )} \nonumber\\
\label{row:ham1}
& - & \mu\sum_i\hat{n}_i - B\sum_i{\hat{s}^z_i},
\end{eqnarray}
where  \mbox{$\hat{n}_{i}=\sum_{\sigma}{\hat{n}_{i\sigma}}$},  \mbox{$\hat{\rho}^+_i=(\hat{\rho}^-_i)^\dag=\hat{c}^+_{i\uparrow}\hat{c}^+_{i\downarrow}$}, \mbox{$\hat{n}_{i\sigma}=\hat{c}^{+}_{i\sigma}\hat{c}_{i\sigma}$},
and $\hat{s}^z_i=(1/2)(\hat{n}_{i\uparrow}-\hat{n}_{i\downarrow})$ is $z$-component of the total spin at $i$ site.
$\hat{c}^{+}_{i\sigma}$ and $\hat{c}_{i\sigma}$ denote the creation and annihilation operators, respectively, of an electron with spin \mbox{$\sigma=\uparrow,\downarrow$} at the site $i$,
which satisfy standard fermion anticommutation relations:
\begin{equation}
\{ \hat{c}_{i\sigma}, \hat{c}^+_{j\sigma'}\}  =  \delta_{ij}\delta_{\sigma\sigma'}, \quad
\{ \hat{c}_{i\sigma}, \hat{c}_{j\sigma'}\}  =  \{ \hat{c}^+_{i\sigma}, \hat{c}^+_{j\sigma'}\} = 0,
\end{equation}
where $\delta_{ij}$ is the Kronecker delta.
$U$ is the on-site density interaction, $I$ is the intersite  charge  exchange interaction (pair hopping) between nearest neighbours,
$B=g\mu_BH_z$ is the external magnetic field, and
$\mu$ is the chemical potential, determining the concentration
of electrons by the formula:
\begin{equation}
n = \frac{1}{N}\sum_{i}{\left\langle \hat{n}_{i} \right\rangle},
\end{equation}
%\mbox{$n = (1/N)\sum_{i}{\left\langle \hat{n}_{i} \right\rangle}$},
with \mbox{$0\leq n \leq 2$}. $\langle\hat{A}\rangle$ denotes the average value of the operator $\hat{A}$ in the grand canonical ensemble,  and $N$ is the total number of lattice sites. \mbox{$\sum_{\langle i,j\rangle}$} indicates the sum over nearest-neighbour sites $i$ and $j$ independently. We also introduce \mbox{$I_0=zI$}, where $z$ is a~number of the nearest-neighbour sites.

It is important to mention that model (\ref{row:ham1}) on the alternate lattices exhibits two symmetries. The first one is a symmetry between \mbox{$I>0$} ($s$-pairing, SS, $\Delta=\frac{1}{N}\sum_i \langle \hat{\rho}^-\rangle$) and \mbox{$I<0$} ($\eta$-pairing, $\eta$S, \mbox{$\Delta_{\eta S} = \frac{1}{N}\sum_i{\exp{(i\vec{Q}\cdot\vec{R}_i)}\langle \hat{\rho}^-_i\rangle} $}, $\vec{Q}$ is half of the smallest reciprocal lattice vector) cases in the absence of the field conjugated with the SS order parameter $\Delta$.
Thus in the following we restrict our analysis to the \mbox{$I>0$} case only. Notice that in the presence of finite single electron hopping \mbox{$t \neq 0$} the symmetry is broken in the general case \cite{HD1993,RB1999,MM2004,PMM2009,PMM2011,RCz2001,CzR2004}.
The second one is the particle-hole symmetry \cite{MRR1990,KRM2012,KR2013}, thus the phase diagrams obtained are symmetric with respect to half-filling and they will be presented only in the range \mbox{$\bar{\mu}=\mu-U/2 \leq 0$} and \mbox{$0\leq n\leq 1$}.

Model (\ref{row:ham1}) has been intensively investigated for \mbox{$B=0$} \cite{B1973,HB1977,WA1987,RP1993,R1994,KRM2012,K2014} as well as for \mbox{$B\neq0$} \cite{RP1993,KR2013} (in particular, in the context of the phase separation \cite{KRM2012,KR2013,N1} (for \mbox{$B=0$}, \mbox{$B\neq0$}) and metastable phases \cite{K2014} (for \mbox{$B=0$})).
The analysis has been performed within a variational approach (VA), which treats the $U$ term
exactly and the intersite $I$ interaction within the mean-field approximation (MFA) (which is a~rigorous treatment of the $I$ term in the limit of infinite dimensions \mbox{$d\rightarrow$} \cite{KRM2012,KR2013,K2014,N1,HB1977}). As a~result for the thermodynamic limit, one gets two
equations for $n$ and $\Delta$, which are solved self-consistently. Equations for the energy and other
thermodynamical properties are derived explicitly in Refs. \cite{KRM2012,KR2013,RP1993,N1}.
\mbox{$|\Delta|\neq 0$} in the superconducting (SS) phase, whereas in the
nonordered (NO) phase \mbox{$|\Delta| = 0$}. For fixed $n$, model (\ref{row:ham1}) can exhibit also the phase separation (PS: SS/NO), which is a~state with two coexisting domains (SS and NO) with different electron concentrations, $n_-$ and $n_+$. The free energy of the PS state is derived in a~standard way, using Maxwell's construction (e.g.~\cite{KRM2012,KR2013,B2004,N1}). It is important to find  homogeneous solutions
for all local minima (even very low ones) with respect to $|\Delta|$ of grand canonical potential $\omega(\mu)$ (or free energy $f(n)$) if system is considered for fixed $\mu$ (or $n$).
The solution (of the set of two self-consistent equations for $n$ and $\Delta$) is related to a~metastable phase if it corresponds to a~(local) minimum of $\omega$ (or $f$) with respect to $|\Delta|$ and the stability condition \mbox{$\partial \mu/\partial n > 0$} (system with fixed $n$) is fulfilled.
Otherwise, the phase is unstable. A~stable (homogeneous) phase is a~metastable phase with the lowest free energy (among  all metastable phases and phase separated states).

%===================================================================================================

\section{Results and discussion}\label{sec:results}

\begin{figure}
    \centering
        \includegraphics[width=\wymiar]{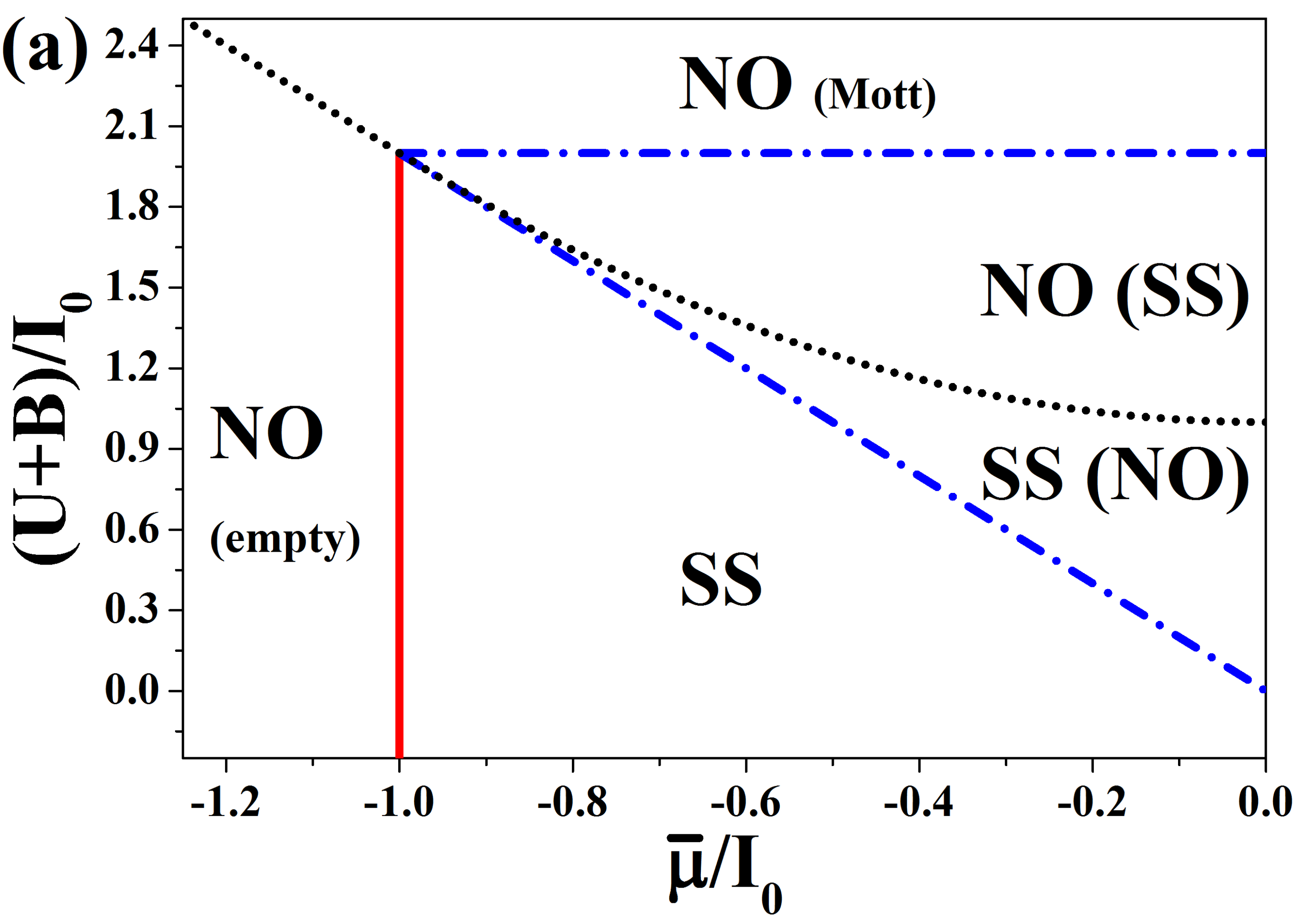}
        \includegraphics[width=\wymiar]{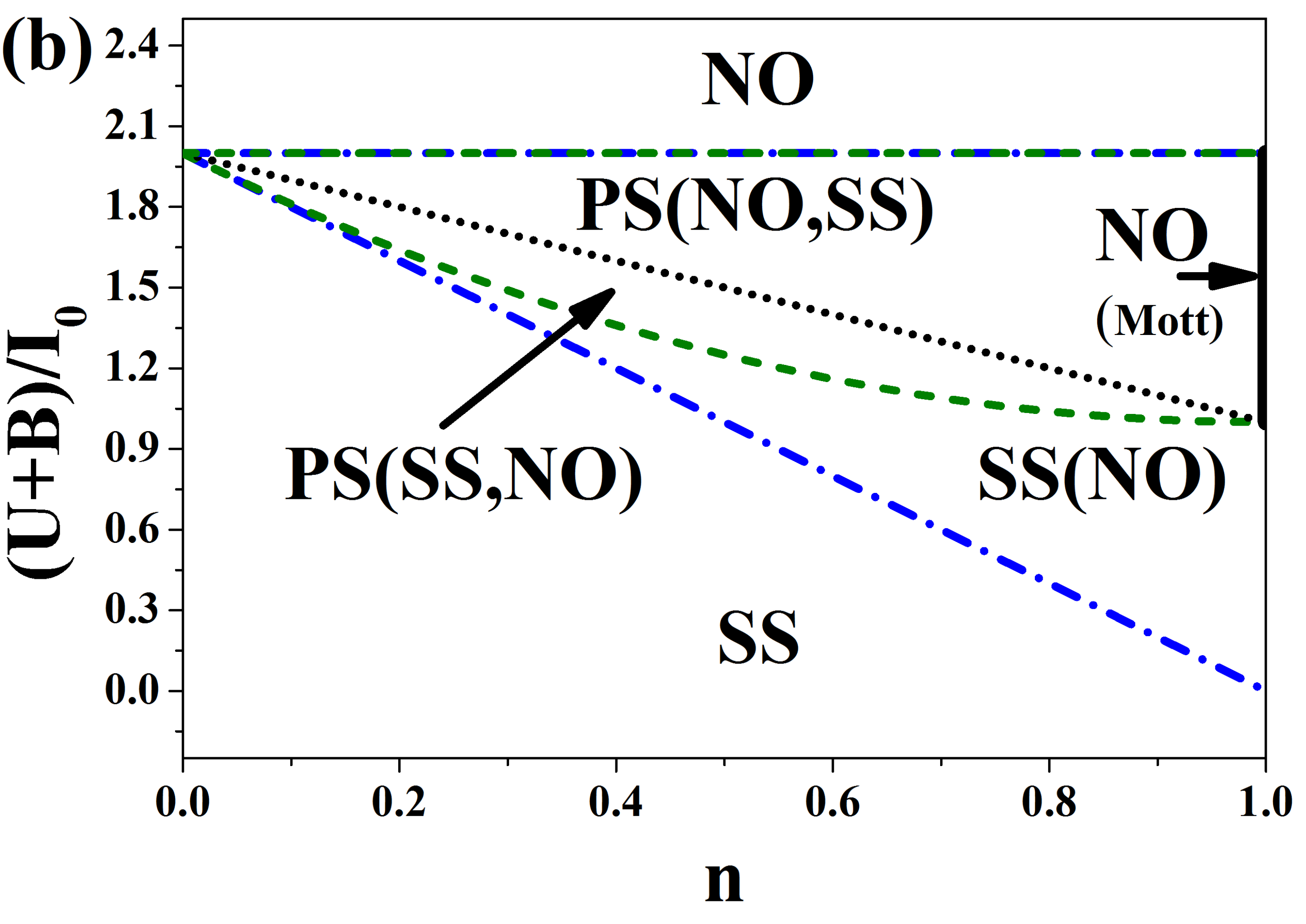}
    \caption{The ground state phase diagrams as a function of $\bar{\mu}/I_0$ (a) and $n$ (b) (\mbox{$\bar{\mu}=\mu-U/2$, \mbox{$I_0=zI$}}). Details in text in Sec.~\ref{sec:results:gs} (cf. also Fig.~\ref{rys:kBTvsmin}).}
    \label{rys:GS}
\end{figure}

Let us distinguish six regions which can occur on the phase diagrams: (1)~only the NO phase is stable; (2)~only the SS phase is stable; (3)~NO(SS) -- in the region of the NO phase stability the SS phase is metastable; (4)~SS(NO) --  the NO phase is metastable in the region of the SS phase stability; (5)~PS(NO,SS) -- the PS state has a lowest energy, the both homogeneous phases are metastable, and the SS phase has a~higher energy than the NO phase; (6)~PS(SS,NO) -- the same as in region (5), but here the NO phase has a~higher energy than the SS phase. Above denotations are used (interchangeably) in Figs.~\ref{rys:GS}--\ref{rys:PSHvskT}.

\subsection{The ground state ($k_BT/I_0=0$)}\label{sec:results:gs}

The ground state (GS) phase diagrams are presented in Fig.~\ref{rys:GS}.
Notice that metastable phases can occur only at \mbox{$k_BT>0$} and the boundaries in Fig.~\ref{rys:GS} of the metastable phases occurrence are the extensions from infinitesimally small \mbox{$T>0$}, formally. At \mbox{$T=0$} one phase (state) can be stable only.
At the GS the discontinuous SS--NO transition occurs at \mbox{$(U+B)/I_0=(\bar{\mu}/I_0)^2+1$} (for fixed \mbox{$|\bar{\mu}|/I_0<1$}) whereas the continuous SS--NO transition occurs at \mbox{$|\bar{\mu}|/I_0=1$} and \mbox{$(U+B)/I_0<2$}.
The PS state (SS/NO) stability region is determined by conditions: \mbox{$(U+B)/I_0\leq2$} and \mbox{$|n-1|^2\leq (U+B)/I_0-1$} (\mbox{$n\neq1$}). At \mbox{$n=1$} (\mbox{$\bar{\mu}=0$}) the discontinuous SS--NO transition occur for \mbox{$(U+B)/I_0=1$}.
The extension (to the GS) of the discontinuous transition line between metastable phases (SS and NO) is located at \mbox{$(U+B)/I_0=1+|1-n|$} (for fixed $n$).
The boundaries for the regions of the metastability of homogeneous phases
%at \mbox{$T>0$} in the vicinity of \mbox{$T=0$}
close to the GS
are located at: for the NO phase, \mbox{$(U+B)/I_0=2|\bar{\mu}|/I_0$} and \mbox{$|\bar{\mu}|/I_0<1$} (\mbox{$(U+B)/I_0= 2|n-1|$}, any $n$); for the  SS phase, \mbox{$(U+B)/I_0=2$} and \mbox{$|\bar{\mu}|/I_0<1$} (and any $n$).
Notice that for the  homogeneous SS phase the condition \mbox{$\partial \mu/\partial n > 0$}  is fulfilled at \mbox{$T=0$} (in particular in the ranges of the PS state occurrence), whereas \mbox{$\partial \mu/\partial n = 0$} in the NO phase \cite{KRM2012,KR2013}.
Let us point out that for \mbox{$T=0$} the discontinuous transition between two NO phases with $|n-1|=1$ (empty/full-filled) and $n=1$ (half-filled Mott state) occurs at \mbox{$(U+B)/I_0=2|\bar{\mu}|/I_0$} and \mbox{$|\bar{\mu}|/I_0>1$}, but it does not exist for any \mbox{$k_BT/I_0>0$}. In fact, the homogeneous NO phase for \mbox{$n\neq1$} is degenerated with the PS state in which two domains of the NO phase (with \mbox{$n_{-}=0$} and \mbox{$n_+=1$} for \mbox{$n<1$} or \mbox{$n_{-}=1$} and \mbox{$n_+=2$} \mbox{$n>1$}) exists. This degeneration is removed for any \mbox{$T>0$} and such a~PS state does not exist at \mbox{$T>0$}.

\subsection{Finite temperatures ($k_BT/I_0>0$)}\label{sec:results:finite}

\begin{figure}
    \centering
        \includegraphics[width=\wymiar]{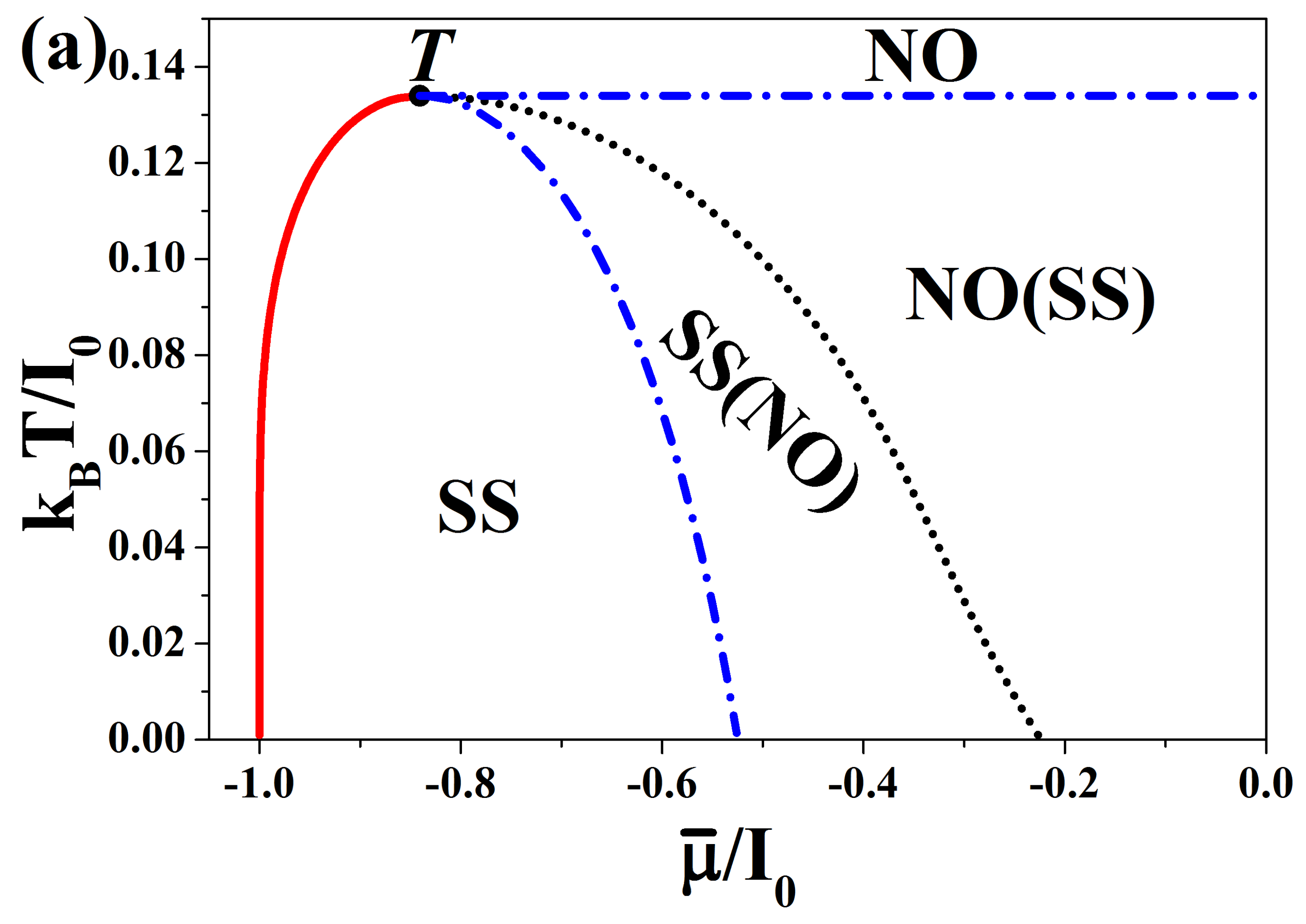}
        \includegraphics[width=\wymiar]{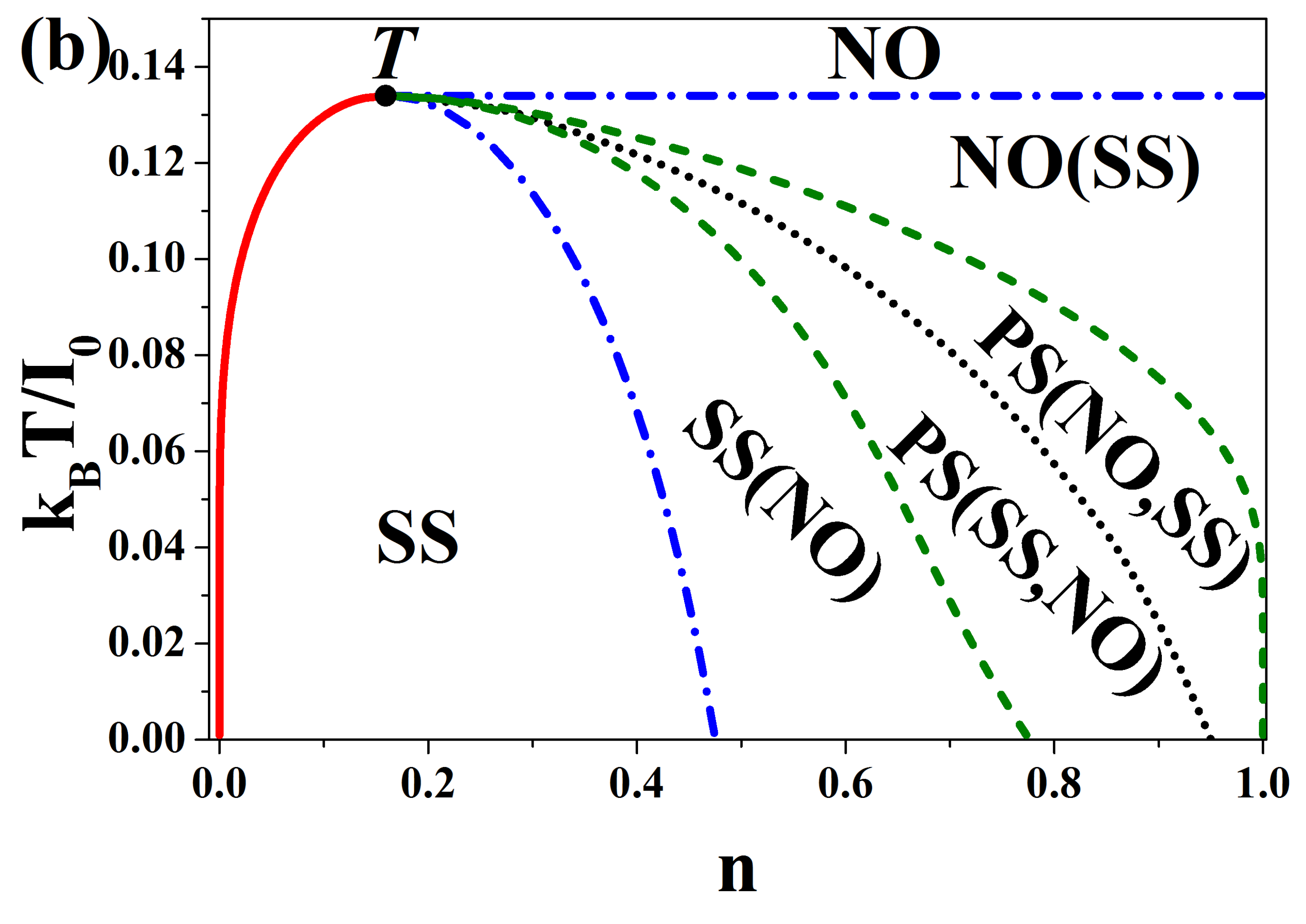}
    \caption{Finite temperature diagrams for \mbox{$U/I_0=1.05$} and \mbox{$B/I_0=0$} as a function of  $\bar{\mu}/I_0$ (a) and $n$ (b). Dotted, solid and dashed lines indicate first-order, second-order and ``third-order'' boundaries, respectively. Dashed-doted lines indicate the boundaries of metastable phase occurrence (names of metastable phases in brackets). $\mathbf{T}$ denotes tricritical point. Details in text in Sec.~\ref{sec:results:finite}.}
    \label{rys:kBTvsmin}
\end{figure}

The complete phase diagram of the model has been determined in \cite{RP1993,KRM2012,KR2013,K2014,N1}. The system analysed shows very interesting multicritical behaviour including tricritical points. Depending on the values of model parameters, the system can exhibit not only the homogeneous phases (SS and NO), but also the phase separated states (PS: SS/NO). \wyroz{All transition temperatures and the SS phase metastability boundary are decreasing functions of $U/I_0$ and $B/I_0$ \cite{KRM2012,KR2013,K2014}. Only the NO phase metastability boundary can exhibit non-monotonic behaviour \cite{K2014}.}

\begin{figure}
    \centering
        \includegraphics[width=\wymiar]{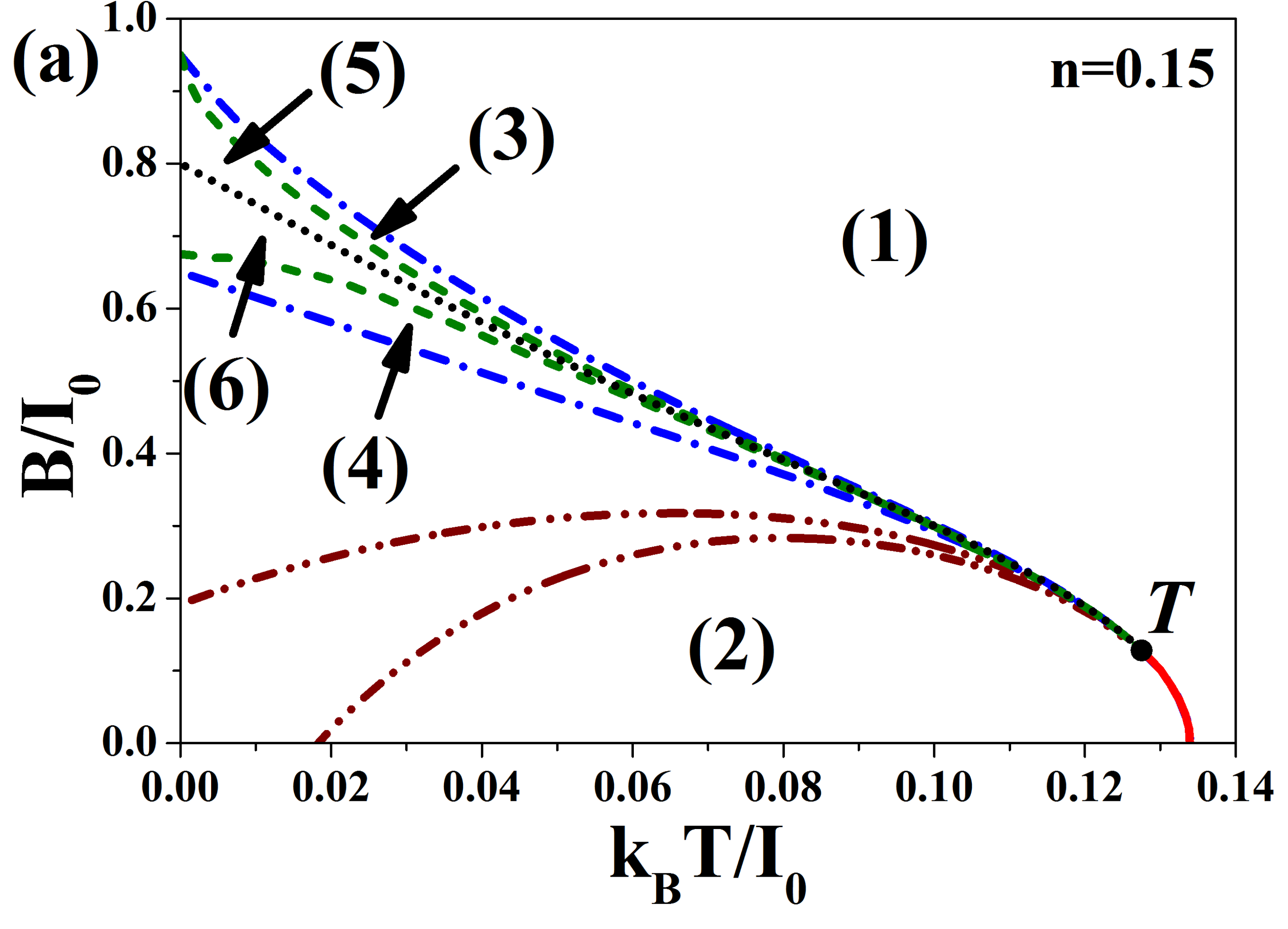}
        \includegraphics[width=\wymiar]{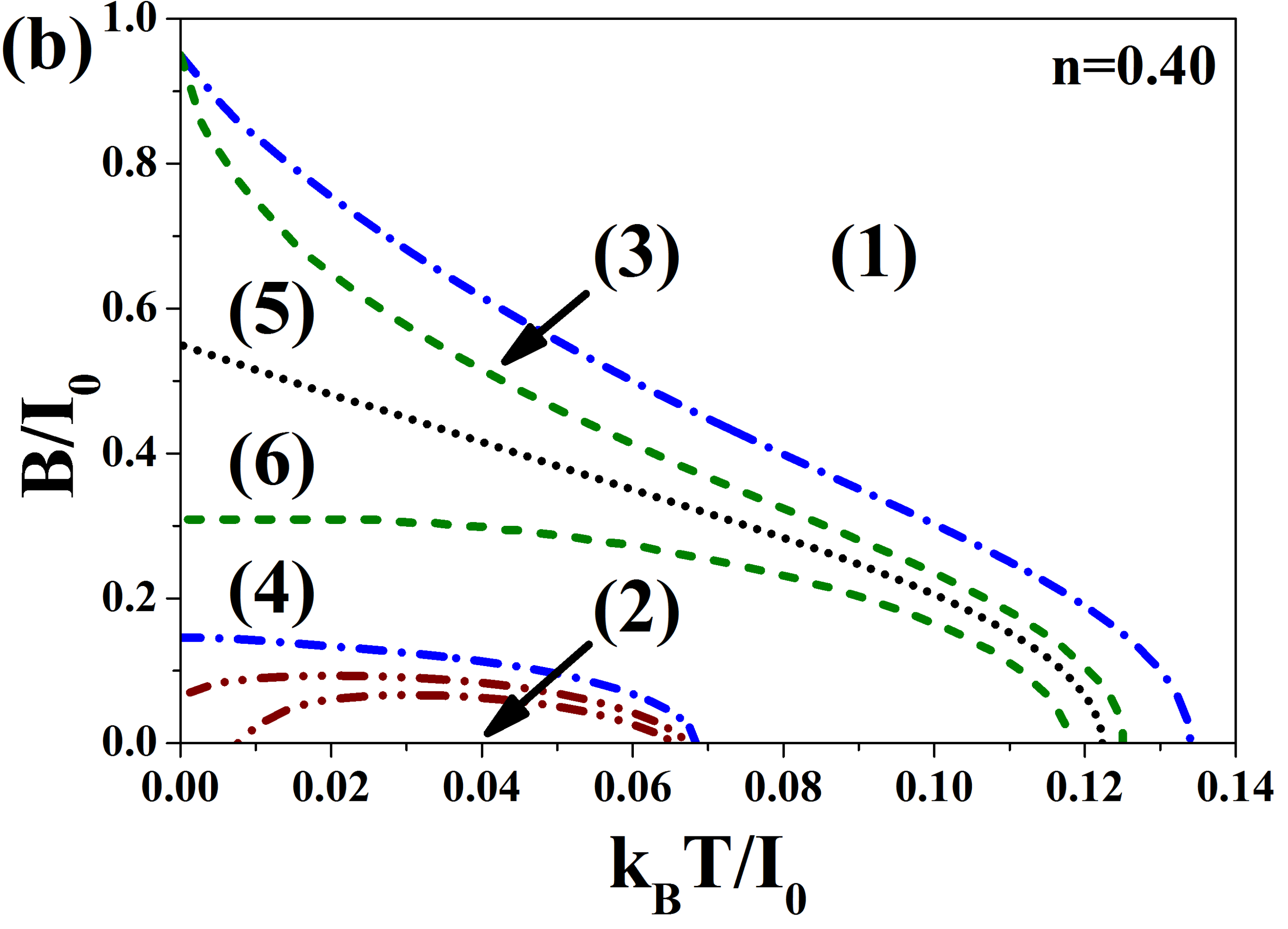}
        \includegraphics[width=\wymiar]{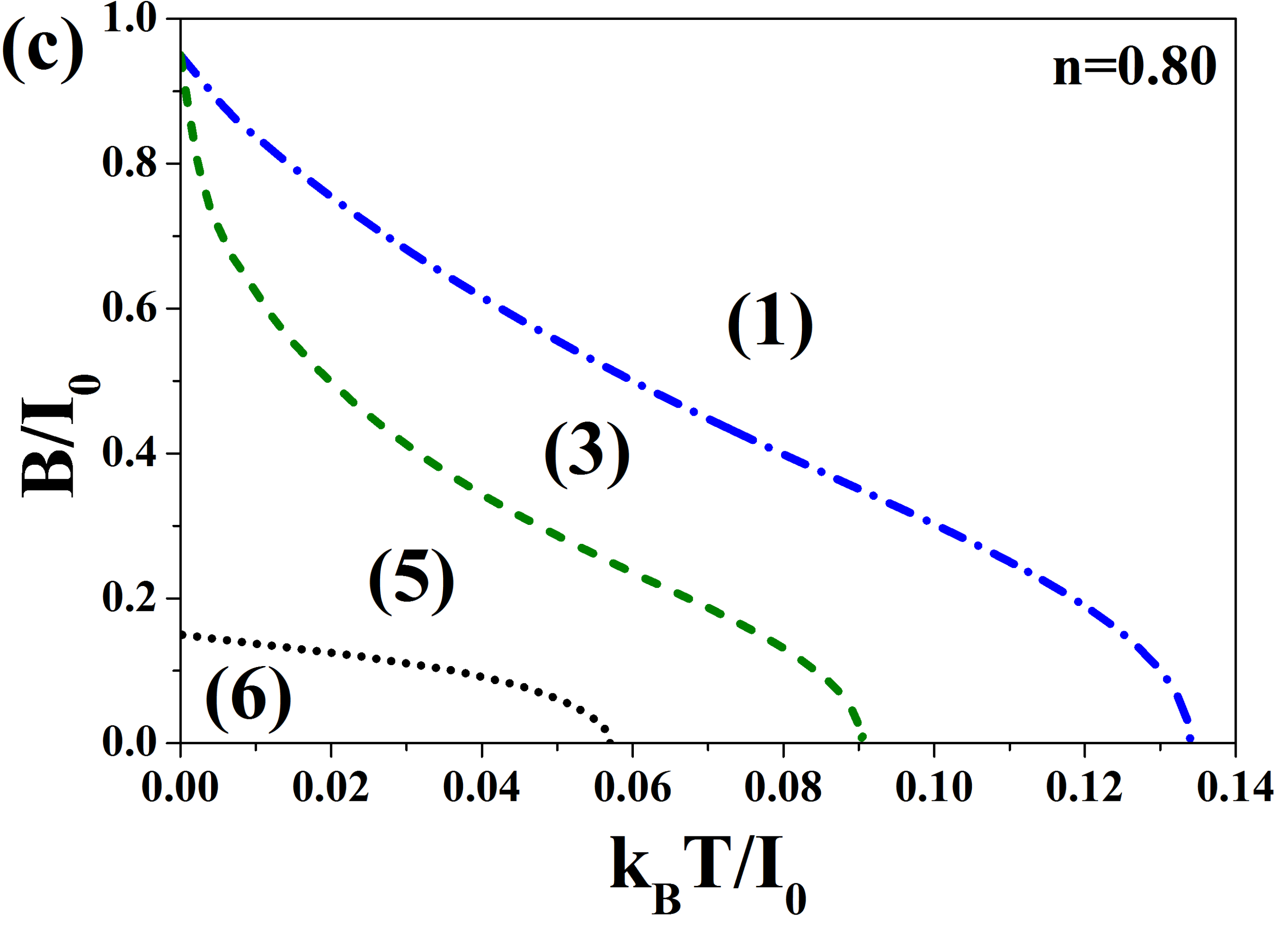}
    \caption{$B/I_0$ vs. $k_BT/I_0$ phase diagrams for  \mbox{$U/I_0=1.05$} and \mbox{$n=0.15,0.40,0.80$} (as labelled). Denotations as in Fig.~\ref{rys:kBTvsmin}. Labels (1)--(6) defined at the beginning of Sec.~\ref{sec:results}. Dashed-dotted-dotted lines (panels (a),(b)) schematically denote  location of the NO phase metastability boundaries (which can occur for other values of $U/I_0$ and $n$ than shown). Details in text in Sec.~\ref{sec:results:finite}.}
    \label{rys:PSHvskT}
\end{figure}

Let us start the discussion of the behaviour of the system for the case \mbox{$1<U/I_0<2$}.
As an example, the phase diagrams for \mbox{$U/I_0=1.05$} and \mbox{$B/I_0=0$} are shown in Fig.~\ref{rys:kBTvsmin}.
The SS--NO transition with increasing temperature can be second-order (continuous change of $\Delta$, the transition temperature decreases with increasing $|\bar{\mu}|/I_0$ and $|n-1|$) as well as first-order (discontinuous change of $\Delta$, the transition temperature increases with increasing $|\bar{\mu}|/I_0$).
It is rather obvious that the regions of the metastable phases occurrence are present
near the first-order \mbox{SS--NO} transition (for fixed $\bar{\mu}$), i.e.
above the  transition temperature the SS phase is metastable (region (3)), whereas below the transition temperature the NO phase is metastable (region (4)), Fig.~\ref{rys:kBTvsmin}(a).
If the system is analysed for fixed $n$, the first-order \mbox{SS--NO} transition line (for fixed $\mu$) splits into two ``third-order'' lines (\mbox{SS--PS} and \mbox{PS--NO}) \cite{KRM2012,KR2013,K2014}. Both ``third-order'' transition temperatures increase with  increasing $|n-1|$ (Fig.~\ref{rys:kBTvsmin}(b)). At this transition a~size of one domain in the PS state decreases continuously to zero at the~transition temperature.
In the region of the PS state occurrence
(where the PS state has the lowest energy $f_{PS}$)
the first-order \mbox{SS--NO} transition between two metastable (homogeneous) phases is present (the transition temperature increase with  increasing $|n-1|$).
Below this line the energy of the NO phase is the highest (i.e. \mbox{$f_{NO}>f_{SS}>f_{PS}$}, region (6)),
whereas above the line the energy of the SS phase is the highest (i.e. \mbox{$f_{SS}>f_{NO}>f_{PS}$}, region (5)), cf. also. Fig.~\ref{rys:PSHvskT}.
The line of the \mbox{SS--NO} first order transition between stable phases for fixed $\bar{\mu}$ (metastable phases for fixed $n$)  ends  at \mbox{$T=0$} and \mbox{$\bar{\mu}<0$} (\mbox{$n<1$}).
One metastable phase (SS or NO) can also be present in the regions of homogeneous phases (NO or SS, respectively) stability for fixed $n$ (where the PS state does not exist), Fig.~\ref{rys:kBTvsmin}(b).

Let us discussed the behaviour of the system with increasing \mbox{$n\leq 1$} for \mbox{$U/I_0=1.05$}.
In Fig.~\ref{rys:PSHvskT} we present a few exemplary $B/I_0$ vs. $k_BT/I_0$ phase diagrams obtained for \mbox{$U/I_0=1.05$} and fixed $n$ (\mbox{$n=0.15,\ 0.40,\  0.80$}).
One can distinguish five essentially different cases.
\begin{itemize}
\item[(i)]~For small $n$ (\mbox{$0<n<0.159$}), the tricritical $\mathbf{T}$-point (associated with a~change of the transition order) is present on the phase diagram (Fig.~\ref{rys:PSHvskT}(a)). With increasing $n$ the $k_BT/I_0$-coordinate of the $\mathbf{T}$-point increases, whereas its $B/I_0$-coordinate  decreases.  At \mbox{$n=0.159$} the $\mathbf{T}$-point is located at \mbox{$B=0$}.
\item[(ii)]~For higher $n$ (\mbox{$0.159<n<0.475$}) all five boundaries remaining on the phase diagrams (i.e. two of ``third-order'', one of first-order between metastable phases, and two of metastable phase occurrence) end at \mbox{$B/I_0=0$} and do not have any points in common (Fig.~\ref{rys:PSHvskT}(b)). These five boundaries  move towards lower values of $B/I_0$ and $k_BT/I_0$ with increasing $n$ and finally three of them  vanish continuously at \mbox{$(k_BT/I_0,B/I_0)=(0,0)$}, whereas two other lines (i.e. the PS--NO ``third-order'' line and the SS phase metastability boundary) are fixed at \mbox{$B/I_0=0.95$} for \mbox{$T=0$}. As the first one, the boundary of the NO phase metastability vanishes at \mbox{$n=0.475$}.
\item[(iii)]~For \mbox{$0.475<n<0.776$}  there is not any region, where the NO phase is unstable (region (2) does not occur on the diagram).
\item[(iv)]~For \mbox{$0.776<n<0.950$} the region of the SS phase occurrence does not exist and for low temperatures only the PS state is stable (the SS phase can be metastable only, region (4) is not present on the diagram, Fig.~\ref{rys:PSHvskT}(c)).
\item[(v)]~At \mbox{$n=0.950$} the boundary of the SS--NO first-order transition between metastable phase disappears and for \mbox{$0.950<n<1$}
below the \mbox{PS--NO} line only region (5) occurs, which shrinks towards lower $T$ with increasing $n$.
\end{itemize}
At \mbox{$n=1$} the PS state does not occur, there is no transitions with increasing $k_BT/I_0$ (the NO phase has always the lowest energy) and the SS phase is metastable in a certain range of model parameters $B/I_0$ and $k_BT/I_0$.

Notice that the boundary of the SS phase metastability is  independent of $n$ (and $\bar{\mu}$) and this line is a projection of the tricritical line (on the $B/I_0$--$k_BT/I_0$ plane in the case of the $B/I_0$ vs. $k_BT/I_0$ diagrams). For small $n$, in the presence of $\mathbf{T}$-point, the boundary does not exist for low $B/I_0$ (Fig.~\ref{rys:PSHvskT}(a)).

For smaller values of  \mbox{$U/I_0<1$} not all behaviours (i)--(v) discussed above and  presented in Figs.~\ref{rys:PSHvskT}(a-c) occur and four new types of phase diagrams can appear.
Two of them are (i'') and (ii''), which structures are similar to these of (i) and (ii), respectively. The  only difference is that the both ends of the NO phase metastability boundary  are located at \mbox{$T>0$} (in particular, the end at lower $T$ is for \mbox{$B=0$}) and the NO phase is unstable (region (2)) only at some \mbox{$T>0$} (cf. lower dashed-dotted-dotted lines in Figs.~\ref{rys:PSHvskT}(a,b)).
Moreover, the boundary of the NO phase metastability can be non-monotonic %with increasing $k_BT/I_0$
(as in the cases (i'') and (ii'')) and can end at \mbox{$T=0$} and \mbox{$B>0$} (as in the cases (i) and (ii)). Thus two more new types: (i') and (ii') can occur, cf. upper dashed-dotted-dotted lines in Fig.~\ref{rys:PSHvskT}(a,b). In general, there is a~continuous change from (i) and (ii) cases to (i'') and (ii'') cases, respectively.

In particular, the following sequences of structures of diagrams with increasing \mbox{$n<1$} occur (some of them in very narrow ranges of model parameters):
\begin{itemize}
\item[(a)] for \mbox{$U/I_0 < 0$}:  only case (i);
\item[(b)] for \mbox{$0< U/I_0 < 0.462$}: (i), (i'), and (i'')
(because the SS--NO transition is always second-order at \mbox{$B=0$} for \mbox{$U/I_0<\frac{2}{3}\ln2$}, cf. Fig.~\ref{rys:PSHvskT}(a));
\item[(c)] for  \mbox{$0.462 <U/I_0 <0.482$}:               (i), (i'), (i''), and (ii'')
\wyroz{(for \mbox{$U/I_0>0.462$} the $\mathbf{T}$-point can occur at \mbox{$B=0$});}
\item[(d)] for  \mbox{$0.482 <U/I_0 <0.557$}:               (i), (i'), (ii'), and (ii'')
\wyroz{(the $\mathbf{T}$-point  at \mbox{$B=0$} moves toward lower $n$ with increasing \mbox{$U/I_0>\frac{2}{3}\ln2$} \cite{KRM2012})}
\item[(e)] for  \mbox{$0.557 <U/I_0 <0.566$}:               (i), (i'), (ii'), (ii''), and (iii)
\wyroz{(for \mbox{$U/I_0>0.557$} the NO phase is metastable for any $T$ near \mbox{$n=1$} for \mbox{$B=0$} \cite{K2014} and case (iii) appears);}
\item[(f)] for  \mbox{$0.566 <U/I_0 <0.666$}:               (i), (ii), (ii'), (ii''), and (iii);
\item[(g)] for  \mbox{$0.666 <U/I_0 <1$}:                   (i), (ii), (ii'), and (iii)
\wyroz{(no re-entrance of region (2) for \mbox{$B=0$} with increasing $T$, the boundary of the NO phase metastability is monotonic at \mbox{$B=0$} for \mbox{$U/I_0>0.666$} \cite{K2014}).}
\end{itemize}

On the contrary to the case \mbox{$1<U/I_0<2$} discussed previously, in above cases (a)--(g)  ``third-order'' \mbox{PS--SS} and  first-order \mbox{SS--NO} (metastable) boundaries connect together at \mbox{$n=1$}. For \mbox{$n=1$} and \mbox{$k_BT/I_0<1/3$}  the first-order SS--NO transition between stable phases is present (cf. Fig.~5 in \cite{KR2013}) and the metastable phases exists in the neighbourhood of this transition (cf. also Fig.~1 in \cite{K2014}). At \mbox{$n=1$} the PS state does not occur.
The NO metastability boundary vanishes continuously to a~point at \mbox{$T=0$} (\mbox{$B=0$}) (the change from (ii') to (iii)) or at \mbox{$T>0$} (\mbox{$B=0$}) (the change from (ii'') to (iii)).

For \mbox{$U/I_0>2$} only the NO phase is stable. There is no metastable phases and no transitions with increasing $k_BT/I_0$.

%\wyroz{%
Notice that determined regions of metastable phases occurrence can be smaller than they actually are due to finite numerical accuracy of local minima finding. This issue can be crucial for determination of (ii') case occurrence, in particular for \mbox{$0.666<U/I_0<2$}.
%}

%===================================================================================================

\section{Final remarks}\label{sec:finalremarks}

The superconductivity with extremely short coherence length and the phase separation  phenomenon involving SS states are very current topics (for a review see \cite{MRR1990,AAS2010,DHM2001,KRM2012,KR2013,K2014,N1} and references therein). It is worthwhile to notice that metastable and unstable  states as well as phase separation have been found in many physical systems experimentally and theoretically.
Note that the temperature dependence of the upper critical field in unconventional superconductors has a positive curvature in coincidence with results of Fig.~\ref{rys:PSHvskT} (cf. PS--NO ``third-order'' line). Obviously the macroscopic PS state founded is different from the Abrikosov-Shubnikov mixed state in type-II superconductors \cite{KR2013}, e.g. no vortex lattice, no magnetic flux quantization, etc.

The results presented in this paper are an extension of our previous investigations of model (\ref{row:ham1}) to the case of \mbox{$B\neq0$} involving the consideration of metastable phases and phase separation.
Model (\ref{row:ham1}) can be considered as a~relatively simple, effective model of a superconductor with local electron pairing \cite{RP1993,KRM2012,K2014,KR2013,MRR1990}. Moreover, the knowledge of the exact \mbox{$d\rightarrow+\infty$} results for the \mbox{$t=0$} limit of the PKH model can be used as a~starting point for a~perturbation expansion in powers of the hopping $t$ and provides a~benchmark for various
approximate approaches analysing the corresponding finite
bandwidth models.

In the model considered the external magnetic field only acts on the spin through the Zeeman term (the paramagnetic effect).
At arbitrary \mbox{$T>0$} the system is spin polarized with non-zero magnetization if \mbox{$B\neq0$}.
A~curious issue  of the orbital contribution through the pair hopping (the diamagnetic effect) \cite{MM2004,RCz2001,R1994} is left for future investigations.
Nevertheless, in materials with heavy electron mass (narrow bands) or multiple small Fermi pockets the paramagnetic  effect becomes crucial. For interacting fermions on non-rotating optical lattices also only paramagnetic effect can occur.

The interplay and competition between superconductivity and intersite magnetic \cite{KKR2010,MKPR2012,MPS2013,MKPR2014,N3} or density-density \cite{MRC1984,MM2008,KR2010,KR2011a,KR2011b,KR2012,MRR1990,MRR1988,N4} interactions  is a~very interesting problem.
Some results concerning the interplay of these interactions with the pair hopping term for \mbox{$B=0$} have been presented in \cite{RP1996,RP1997,K2012,K2013,CzR2006,N1}.

%===================================================================================================

\begin{acknowledgements}
The author is indebted to Professor Stanis\l{}aw Robaszkiewicz for very fruitful discussions during this work and careful reading of the manuscript.
The work has been financed by National Science Centre (NCN, Poland) as a~research project under grant No. DEC-2011/01/N/ST3/00413 and a~doctoral scholarship No.  DEC-2013/08/T/ST3/00012.
The author also thanks the European Commission and the Ministry of Science and Higher Education (Poland)
for the partial financial support from the European Social Fund---Operational Programme ``Human Capital''---POKL.04.01.01-00-133/09-00---``\textit{Proinnowacyjne kszta\l{}cenie, kompetentna kadra, absolwenci przysz\l{}o\'sci}''
as well as the Foundation of Adam Mickiewicz University in Pozna\'n for the support from its scholarship programme.
\end{acknowledgements}

%===================================================================================================

\end{document}